\pdfoutput=1

\documentclass[11pt]{article}

\usepackage[final]{acl}

\usepackage{times}
\usepackage{latexsym}

\usepackage[T1]{fontenc}

\usepackage[utf8]{inputenc}

\usepackage{microtype}

\usepackage{inconsolata}

\usepackage{graphicx}
\usepackage{adjustbox}
\usepackage{xcolor}
\usepackage{multirow}
\usepackage{booktabs}
\usepackage{url}
\usepackage[colorinlistoftodos]{todonotes}
\usepackage{amssymb}
\usepackage{xspace}

\newcommand{\method}{\textit{SYN2REAL}\xspace}
\newcommand{\methodd}{\textit{SYN2REAL Ensemble}\xspace}

\title{Task Arithmetic can Mitigate Synthetic-to-Real Gap in \\ Automatic Speech Recognition}

\author{
    Hsuan Su$^\heartsuit$ \qquad
    Hua Farn$^\heartsuit$ \qquad
    Fan-Yun Sun$^\diamondsuit$ \qquad
    Shang-Tse Chen$^\heartsuit$ \qquad
    Hung-yi Lee$^\heartsuit$
    \\
    $^\heartsuit$National Taiwan University \qquad
    $^\diamondsuit$Stanford University \\
   \small \texttt{hsuansu.96@gmail.com}
}

\begin{document}
\maketitle
\begin{abstract}
Synthetic data is widely used in speech recognition due to the availability of text-to-speech models, which facilitate adapting models to previously unseen text domains. However, existing methods suffer in performance when they fine-tune an automatic speech recognition (ASR) model on synthetic data as they suffer from the distributional shift commonly referred to as the synthetic-to-real gap. In this paper, we find that task arithmetic is effective at mitigating this gap. Our proposed method, \method task vector, shows an average improvement of 10.03\% improvement in word error rate over baselines on the SLURP dataset. Additionally, we show that an average of \method task vectors, when we have real speeches from multiple different domains, can further adapt the original ASR model to perform better on the target text domain.


\end{abstract}
\section{Introduction}
\label{sec:intro}

Existing automatic speech recognition (ASR) models have been found to lack generalizability towards domains unseen during training \cite{bartelds2023making, radford2022robust, Sundar2023MultimodalAM}. Existing works, when adapting an ASR model to a previously unseen domain, often rely on synthetic speech data \cite{corpus-synthesis, bataev2023textonly, joshi-singh-2022-simple, 9414778, 10317116, 10096971} due to its ease of generation and availability. However, this approach often leads to performance degradation due to acoustic mismatches such as intonations, background noise, speaker accents, and environmental sound differences between synthetic and real speech~\cite{corpus-synthesis}. This distributional shift is often referred to as the synthetic-to-real gap. This paper tackles this problem, particularly when adapting an ASR model from a source domain with text and real speech data to a new target domain with only text data.

Our idea is that paired synthetic speech and real speech data within a single domain can guide the adaptation of models trained on synthetic data to perform better on real-world data in a new domain. Inspired by the new paradigm of editing pre-trained neural networks by manipulating their weights \cite{sung-etal-2023-empirical, tam2024merging}, we propose to bridge the synthetic-to-real gap using task vectors~\cite{chatvector, task_vector-homer-bhardwaj2024language}. A task vector is a representation that encodes the difference between two tasks, allowing models to perform arithmetic operations to transition from one task to another. In this paper, we show that we can apply simple arithmetic operations to bridge the synthetic-to-real gap.


\begin{figure}[tp!]
    \begin{adjustbox}{width=\linewidth}
    \centering
    \includegraphics{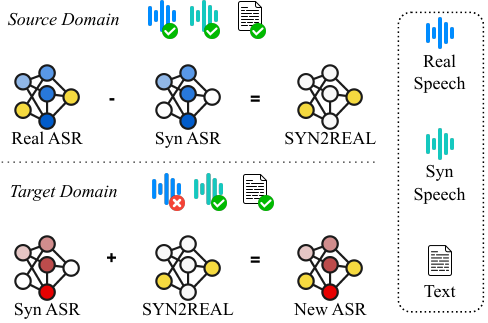}
    \end{adjustbox}
    \caption{\textbf{Overview of the \method{} Task Vector Approach.} The pre-trained model is fine-tuned on source domain synthetic and real speech data, separately. The difference between their parameters forms the \method task vector. The \method task vector is then added to a model fine-tuned on target synthetic data to overcome the synthetic-to-real gap.}
    \vspace{-3mm}
     \label{fig:overview}
\end{figure}


Taking inspiration from~\citet{ilharco2023editing}, we propose \method task vector for synthetic-to-real adaptation in ASR. Figure~\ref{fig:overview} provides an overview of the \method task vector approach. The top row illustrates the process of fine-tuning models on synthetic and real speech data separately and then deriving the \method task vector from the differences in their parameters. The bottom row demonstrates the application of this vector to a model fine-tuned on synthetic target domain data, resulting in an adapted model with improved performance by incorporating the acoustic characteristics of real speech.


We conduct comprehensive experiments and ablation studies to demonstrate the effectiveness of our approach. Applying the \method task vector results in an average improvement of 10.03\% in word error rate (WER) for unseen target domains compared to the model before applying our method. Cosine similarity analysis of \method task vectors generated by different text-to-speech (TTS) models confirms that \method task vectors effectively capture domain-specific acoustic information.


\section{Related Works}
\label{sec:related}
\paragraph{ASR Text-only Domain Adaptation}
In the context of automatic speech recognition (ASR), "text-only" domain adaptation typically refers to scenarios where the target domain only provides text data for training or fine-tuning the models. Previous work has explored internal language models adaptation that finetune language models in ene-to-end ASR models with CTC loss to improve the generalizability \cite{10389682,sato22_interspeech, vuong-etal-2023-adabert}.

The other direction is to adapt ASR models with synthetic speech. \citet{9414778} develop a method that provides synthetic audio for out-of-vocabulary (OOV) words to boost recognition accuracy. \citet{10096971} works on personalizing ASR with synthetic speech. \citet{bataev2023textonly} focuses on developing a mel-spectrogram generator to improve ASR models.


\paragraph{Task Arithmetic}
The concept of task vector is introduced in~\citet{ilharco2023editing}. Task vectors are created by subtracting the weights of a fine-tuned model from those of its corresponding pre-trained model. 
Different task vectors derived from the same pre-trained models can then be adjusted and combined through these simple arithmetic operations such as addition and subtraction to achieve multi-task learning~\cite{task_vector-zhang2023composing} and task forgetting~\cite{task_vector-daheim2023elastic}.

Recently, task vectors have shown promise in natural language processing (NLP) \cite{chatvector,task_vector-daheim2023elastic,task_vector-homer-bhardwaj2024language,task_vector-zhang2023composing}. \citealt{task_vector-daheim2023elastic} used a task vector from a negatively fine-tuned model to mitigate hallucinations. \citet{task_vector-zhang2023composing} proposed combining parameter-efficient fine-tuning (PEFT) modules \cite{lora, ia3} arithmetically. \citet{chatvector} obtained the Chat Vector by subtracting the chat version of Llama 2 \cite{touvron2023llama} from its pre-trained version, enhancing dialogue capabilities and safety. \citet{task_vector-homer-bhardwaj2024language} introduced RESTA, adding a safety vector to re-align safety for models fine-tuned on downstream tasks. The application of task vectors is relatively underexplored in ASR. \citet{10447848} applied task arithmetic to ASR models and introduced a "task analogy" formulation, improving performance on low-resource tasks using models trained on high-resource tasks. Unlike~\citet{10447848}, we focus on using task vector to mitigate the distributional shift between real and synthetic data.

\section{Methodology}
\label{sec:method}

In the context of automatic speech recognition (ASR) domain adaptation, domain mismatch can be broadly classified into three categories:

\begin{enumerate}
    \item \textbf{Acoustic Variation Mismatch}: This mismatch refers to differences in speech caused by variations in acoustic properties.
    \item \textbf{Textual Topic Mismatch}:
        This mismatch involves discrepancies in the subject matter or style of the textual content.
    \item \textbf{Synthetic vs. Real Speech Mismatch}: This mismatch refers to the acoustic differences between synthesized speech generated from text and actual spoken speech.
\end{enumerate}
\begin{figure}[hp!]
    \begin{adjustbox}{width=\linewidth}
    \centering
    \includegraphics{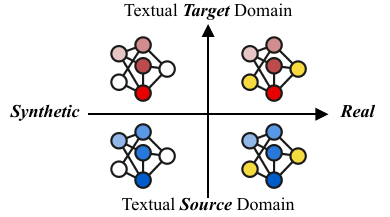}
    \end{adjustbox}
    \caption{\textbf{Domain Shifts in ASR Domain Adaptation.} Illustration of domain adaptation challenges in ASR, showing shifts between synthetic and real speech across source and target textual domains.}
     \label{fig:domain_shift}
\end{figure}
In this work, we aim to adapt ASR models from source textual domains with real speech and text data to a new textual domain with only text data. We leverage data synthesized from off-the-shelf text-to-speech (TTS) systems to address this textual topic mismatch. Figure~\ref{fig:domain_shift} illustrates the domain shifts we focused on in ASR adaptation, depicting the challenges of bridging both the textual gap (source vs. target domain) and the acoustic gap (synthetic vs. real speech).

While previous works \cite{10096971, corpus-synthesis} have shown that adapting ASR models using synthetic data effectively addresses textual topic mismatch, ASR models trained on synthetic data often underperform compared to those trained on real data due to mismatches between synthetic and real speech.

To overcome this limitation, we propose the \method task vector, a novel approach designed to bridge the acoustic gap between synthetic and real speech data, enhancing the performance of ASR models in domain adaptation.

\subsection{Problem Formulation}
We assume a problem setting in which we have two domains: a source domain $D_s$ and a target domain $D_t$. 
The source domain $D_s$ consists of paired text and speech samples, denoted as $T_s$ and $S_s$, respectively. The target domain $D_t$ contains only text data, denoted as $T_t$. This problem setting is common as it is easy to generate synthetic text data, whereas collecting paired real speech data is labor-intensive.

\subsection{\method Task Vector}
To adapt ASR models to a previously unseen domain, we employ a common methodology \cite{joshi-singh-2022-simple} that utilizes synthetic data generated from the target text $T_t$ for model adaptation. 


\begin{figure}[htp!]
\begin{adjustbox}{width=\linewidth}
\centering
\includegraphics{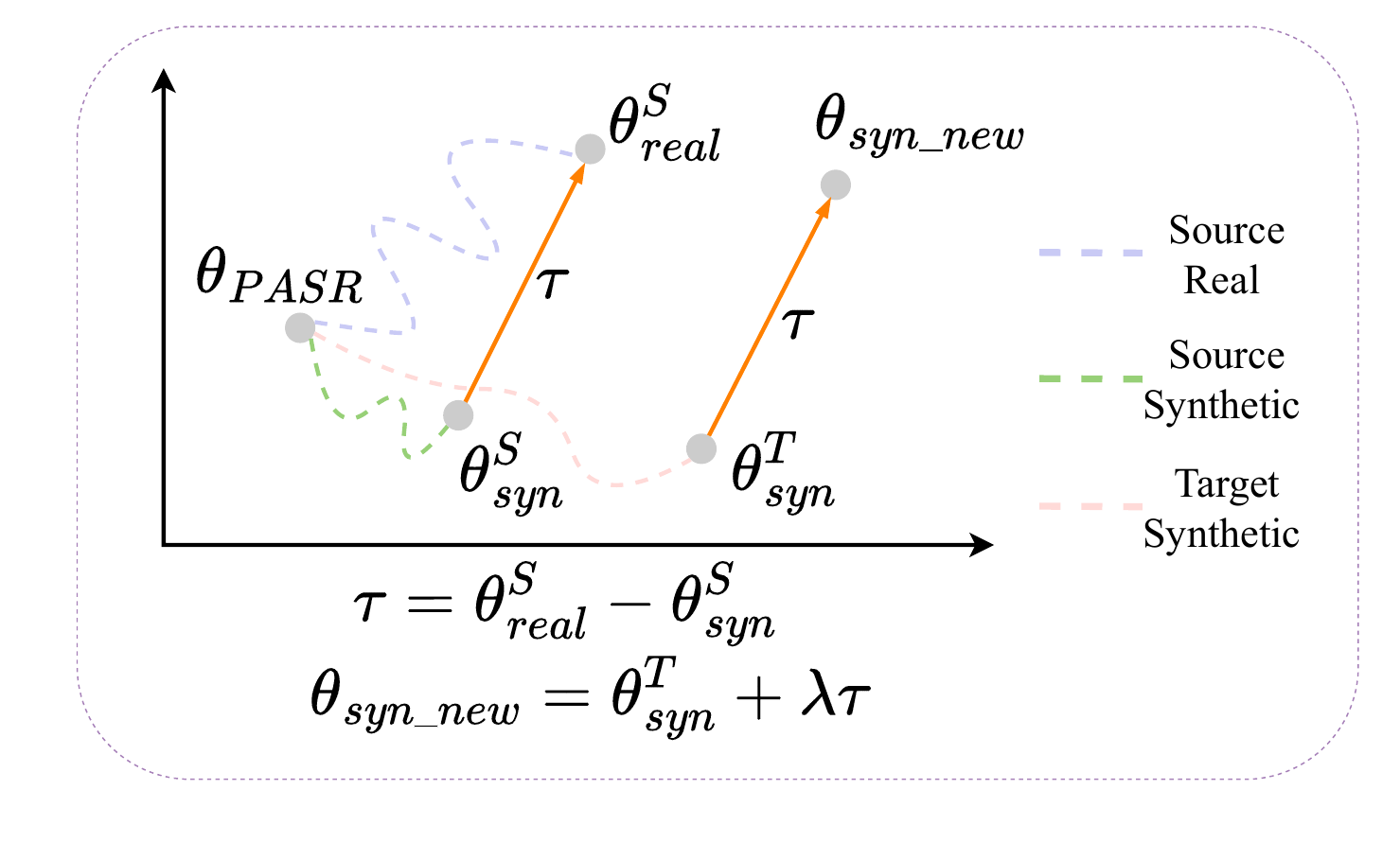}
\end{adjustbox}
\caption{\textbf{Framework for \method task vector in Domain Adaptation for ASR.} The framework illustrates the process of creating the \method task vector by subtracting the parameter differences between a model fine-tuned on synthetic speech (Source Synthetic) and a model fine-tuned on real speech (Source Real) from pretrained ASR (PASR). This task vector is then applied to the target synthetic domain (Target Synthetic) to improve ASR performance by bridging the gap between synthetic and real speech data.}
\label{fig:method}
\end{figure}

Previous work in task arithmetic has demonstrated that vectors can encode distinct capabilities, such as language or domain-specific features. We hypothesize that the differences in acoustic properties between real and synthetic speech are also learnable and can be isolated through parameter arithmetic. Specifically, we assume that we have models fine-tuned on real and synthetic data from the source domain, denoted as $\theta_{real}^{S}$ and $\theta_{syn}^{S}$ respectively.
The acoustic disparity between real and synthetic speech is quantified by subtracting the parameter sets of these models:
\begin{equation}
\tau = \theta_{real}^{S} - \theta_{syn}^{S}
\end{equation}

Once the \method vector $\tau$ is computed, we apply it to the model parameters fine-tuned on the synthetic data in the target domain, $\theta_{syn}^{T}$, thereby enhancing its adaptation to the target domain:
\begin{equation}
\theta_{syn\_new} = \theta_{syn}^{T} + \lambda \tau
\end{equation}
Where $\lambda$ is the scaling factor of \method task vector.

This adjusted model, $\theta_{syn\_new}$, is expected to perform more robustly in the target domain as it incorporates the acoustic characteristics of real speech, making it better suited for practical ASR tasks where real speech is present.

\paragraph{\methodd Task Vector}
In the previous discussion, we assume no access to domain labels such as 'email' or 'music' from the source domain to emulate real-world situations better. All real speech data falls in the source domain. 
In scenarios where we have access to data of multiple domains, another approach is to create \method task vectors for each domain separately and then combine these vectors. This method involves fine-tuning separate ASR models on each individual source domain to obtain domain-specific \method task vectors, which are then averaged to form a comprehensive \method task vector.

For each domain $i$ in source domain $S$. The task vector can be defined as:
\begin{equation}
\tau_i = \theta_{real}^{S_i} - \theta_{syn}^{S_i}
\end{equation}
Where  $\theta_{real}^{S_i}$ and $\theta_{syn}^{S_i}$ represent the model parameters fine-tuned on real and synthetic data for domain $i$ respectively.
Once the vector $\tau_i$ is computed, we apply it to the model parameters fine-tuned on synthetic target domain data $\theta_{syn}^{T}$, thereby enhancing its adaptation to the target domain:
\begin{equation}
\theta_{syn\_new} = \theta_{syn}^{T} + \frac{\lambda}{|S|} \sum_{i=0}^{|S|} \tau_i
\end{equation}
Where $|S|$ is the number of source domains, and $\lambda$ is the scaling factor for the task vector.
\section{Experimental Setups}
We design our experiments to answer the following questions: \textbf{Q1: }What is the efficacy of \method task vector?, \textbf{Q2: }How does \method task vector perform across different model sizes? \textbf{Q3: }Is \method task vector effective on ASR models other than Whisper?, \textbf{Q4: }Can we form \method task vectors from other TTS models?, \textbf{Q5: }What is the impact of the scaling factor $\lambda?$ , \textbf{Q6: }Do \method task vectors obtained with the same TTS have similar direction?

To address these questions and simulate real-world scenarios, we first create a source domain ASR model by combining synthetic and real speech data from various domains. We then adapt this source domain ASR model to the target domain using data synthesized by TTS models.
\method task vector is constructed by subtracting the weights of an ASR model fine-tuned on synthetic data from the weights of the same ASR model fine-tuned on real data, both using the same pre-trained model as the starting point. Our goal is to improve the performance of an ASR model on the target domain without using any real speech from the target domain.
\begin{table*}[t]
\begin{adjustbox}{width=\linewidth}
\begin{tabular}{cccccccccccccccccccc}
\toprule
\textbf{WER}                                                   & \multicolumn{18}{c}{\textbf{Target Domains}}                                                                                        & \multirow{2}{*}{\textbf{Average}} \\ \cmidrule{1-19}
\textbf{Methods}                                                           & \textbf{Alarm}            & \textbf{Audio}            & \textbf{Calendar}         & \textbf{Cooking}          & \textbf{Datetime}         & \textbf{Email}            & \textbf{General}          & \textbf{IOT}              & \textbf{Lists}            & \textbf{Music}            & \textbf{News}             & \textbf{Play}             & \textbf{QA}               & \textbf{Recommendation}   & \textbf{Social}           & \textbf{Takeaway}         & \textbf{Transport}        & \textbf{Weather}          &                          \\ \midrule
\begin{tabular}[c]{@{}c@{}}Target Synthetic ASR\\ (Baseline)\end{tabular} & 16.13              & 14.69             & 22.88             & 14.26             & 47.16             & 16.23             & 27.16             & 13.67             & 15.49             & 23.51             & 21.31             & 21.61             & 24.04             & 17.54             & 29.57             & 21.25             & 18.91             & 15.45             & 21.16                    \\ \midrule
+ \method                                                          & \textbf{15.65}    & \textbf{13.68}    & \textbf{22.64}   & 14.36    & \textbf{40.29}    & \textbf{16.15}   & \textbf{16.87}  & \textbf{12.49}   & \textbf{15.22}   & \textbf{17.03}    & \textbf{21.25}   & \textbf{20.77}   & \textbf{23.88}   & \textbf{15.19}   & \textbf{21.87}   & \textbf{18.03}   & \textbf{16.90}    & 20.38    & \textbf{19.04}           \\
Relative WER (\%)$\uparrow$                                                   & \textbf{2.95\%} &  \textbf{6.87 \%} & \textbf{1.03 \% } & -0.70 \% & \textbf{14.58 \%} & \textbf{0.50 \%} & \textbf{37.89 \%} & \textbf{8.58 \%} & \textbf{1.74 \%} & \textbf{27.57 \%} & \textbf{0.28 \%} & \textbf{3.88 \%} & \textbf{0.64 \%} & \textbf{13.42 \%} & \textbf{26.04 \%} & \textbf{15.14 \%} & \textbf{10.65 \%} & -31.91\%   & \textbf{10.03\%}       \\ \bottomrule

\end{tabular}
\end{adjustbox}
\caption{\textbf{Word Error Rate (WER) Performance Across Various Target Domains.}
Comparison of the baseline Whisper model and the model enhanced with the \method task vector generated by BARK. The \method task vector shows an average WER reduction of 10.03\% across various target domains. Target Synthetic ASR refers to the baseline that is finetuned on 17 domains (excluding the target domain) real+synthetic data followed by synthetic data from the target domains in the SLURP dataset.}
\label{tab:11000table}
\end{table*}
\subsection{Dataset}
SLURP~\cite{bastianelli-etal-2020-slurp} is a spoken language understanding dataset containing 16521 utterances of human commands towards a virtual agent, based on 200 pre-defined prompts such as ``How would you ask for the time.'' The utterances are categorized into 18 domains (e.g., email, cooking, etc.). In each of our experiments, we select one of these domains as the target domain and combine the remaining 17 domains to form the source domain.

\subsection{Text-to-Speech (TTS) Models}
In our experiments, we prepare synthetic speech using two off-the-shelf TTS models for text from the target domains.
\paragraph{\textbf{BARK}} BARK\footnote{\url{https://github.com/suno-ai/bark}} is a transformer-based \cite{vaswani2023attention} autoregressive model, it is pretrained with similar architecture as AudioLM \cite{borsos2023audiolm} and Vall-E \cite{wang2023neural}. The input of BARK contains prompts, transcription, and users. In our experiments, we do not specify the speaker for BARK.
  
\paragraph{\textbf{Speech T5}} 
Speech T5~\cite{ao-etal-2022-speecht5} is a unified model framework that employs encoder-decoder pre-training for self-supervised speech/text representation learning. 
In our experiments, we randomly sample 5 speakers from 7931 speakers.

\paragraph{\textbf{XTTS}}
XTTS is a SOTA TTS model released by Coqui \cite{casanova2024xttsmassivelymultilingualzeroshot}. It is a multi-speaker, end-to-end TTS model capable of synthesizing production-level quality speech. In our experiments, we use its second version, XTTS-v2\footnote{\url{https://huggingface.co/coqui/XTTS-v2}}, to synthesize speech for the target domain.

\subsection{ASR Models} 
\paragraph{Whisper} 
Whisper \cite{radford2022robust} is an encoder-decoder Transformer-based \cite{vaswani2023attention} model that supervised finetuned on 680,000 hours of labeled audio data.
All experiments are conducted using the Whisper small model, except for the ablation study, where we experiment with models of different sizes, including the base and tiny models, to validate our method.
\paragraph{Wav2Vec2-Conformer} 
Wav2Vec2 \cite{baevski2020wav2vec} is a framework for self-supervised learning of speech representations that masks latent representations of the raw waveform and solves a contrastive task over quantized speech representations.
Wav2Vec2-Conformer \cite{wang2022fairseq} (referred to as Wav2Vec2 in the experiments) follows the same architecture as Wav2Vec2, but replaces the Attention-block with a Conformer-block \cite{wang-etal-2020-fairseq} is the conformer \cite{gulati20_interspeech}. We use the large checkpoint\footnote{\href{https://huggingface.co/facebook/wav2vec2-conformer-rope-large-960h-ft}{facebook/wav2vec2-conformer-rope-large-960h-ft}} with 618M parameters with rotary position embeddings, pretrained and fine-tuned on 960 hours of Librispeech \cite{7178964} on 16kHz sampled speech audio to conduct experiments.


\section{Results \& Discussion}
Here, we discuss our results in relation to the questions we set out to answer.

\subsection{What is the efficacy of \method task vector?}
To answer \textbf{Q1}, we apply our method by comparing the word error rate (WER) across various target domains. We select one of these domains as the target domain and combine the remaining 17 domains to form the source domain
Table~\ref{tab:11000table} presents the WER results for both the baseline ASR model fine-tuned on synthetic speech data and the model enhanced with the \method task vector.

The baseline model, fine-tuned solely on synthetic data, exhibits varying WERs across different target domains, with an average WER of 21.16. This performance highlights the challenge of adapting ASR models to real-world data when trained on synthetic speech, primarily due to acoustic mismatches.

The application of the \method task vector significantly reduces WER across most target domains.
The \method-enhanced model achieves an average WER of 19.04, representing an average relative WER reduction of 10.03\%. This improvement demonstrates the \method task vector's effectiveness in bridging the gap between synthetic and real speech data, enhancing the model's adaptability to diverse real-world scenarios.

To further validate our method's effectiveness in solving the synthetic vs. real speech mismatch, we also report two additional baselines. We conducted experiments evaluating the pretrained Whisper small model and the Whisper small model fine-tuned on the source domain (real + synthetic) on the target domain, with average WERs of 33.30 and 27.13, respectively. We address the acoustic variation mismatch by fine-tuning the pretrained Whisper on source domain speech. Further fine-tuning on target domain synthetic data addresses the textual topic mismatch, achieving a WER of 21.16. Finally, applying the \method task vector mitigates the synthetic vs. real speech mismatch, resulting in a WER of 19.13.

The \method task vector shows particularly notable improvements in domains such as 'Music' (27.57\% reduction), 'Takeaway' (15.14\% reduction), and 'Social' (26.04\% reduction). These results suggest that the task vector effectively captures domain-specific acoustic variations, enabling the ASR model to generalize better to unseen target domains. In the following experiments we select the four domains includes two highest improved domains ('Music' \& 'Social'), and the two lowest improved domains ('Weather' \& 'Cooking') to conduct the experiments. 

However, it is important to note that some domains, such as 'Cooking' and 'Weather,' exhibit marginal improvements or slight degradation in WER. These variations indicate that while the \method task vector generally enhances performance, further fine-tuning and domain-specific adjustments may be necessary to optimize results across all target domains.

Overall, the results demonstrate that the \method task vector is a promising approach for improving ASR domain adaptation, showing significant improvements over both the baseline and a recent work, AdaBERT-CTC \cite{vuong-etal-2023-adabert}, which only achieved a WER of 26.1 on the SLURP test set. By addressing the acoustic mismatches between synthetic and real speech data, our method significantly enhances the performance of ASR models in real-world applications.
\begin{table}[ht!]
\begin{adjustbox}{width=\linewidth}

\begin{tabular}{cccccc}
\toprule
\textbf{Relative WER} $\uparrow$ & \textbf{Cooking} & \textbf{Music}   & \textbf{Social}  & \textbf{Weather}  & \textbf{Average}          \\ \midrule
Tiny         & \textbf{41.11\%}  & -13.47\% & 2.60\%  & \textbf{30.42\%}  & \textbf{19.48\%}  \\ \midrule
Base         & 1.49\%  & \textbf{37.80\%} & 5.00\%  & 6.82\%   & 14.70\% \\ \midrule
Small        & -0.70\% & 27.56\% & \textbf{26.04\%} & -31.91\% & 12.43\% \\ \bottomrule
\end{tabular}

\end{adjustbox}
\caption{\textbf{Relative WER Improvement Across Different Model Sizes after applying \method task vector.} This table shows the relative WER improvement compared to the Target Synthetic ASR for Whisper models of various sizes (Tiny, Base, and Small).}
\label{tab:size}
\end{table}
\subsection{How does \method task vector perform across different model sizes?}

To answer \textbf{Q2}, we analyze the effect of model size on the performance of ASR adaptation using the \method task vector. Table~\ref{tab:size} presents the relative word error rate improvements across different model sizes (Tiny, Base, Small) and various target domains.

The results indicate that the Base model achieves the highest average relative WER improvement of 14.70\% across all target domains. This model size shows substantial gains, particularly in the 'Music' (37.80\%) and 'Social' (5.00\%) domains, demonstrating its robustness in adapting to diverse acoustic characteristics using the \method task vector.

The Tiny model, while achieving a higher average improvement of 19.48\%, shows considerable performance gains in the 'Cooking' (41.11\%) and 'Weather' (30.42\%) domains. However, it experiences a performance degradation in the 'Music' domain (-13.47\%). This suggests that while the Tiny model can benefit significantly from the \method task vector in certain domains, its overall adaptability might be limited compared to larger models due to its reduced model size.

Interestingly, the Small model exhibits an average relative WER improvement of 12.43\%, with substantial gains in the 'Social' (26.04\%) and 'Music' (27.56\%) domains. However, it shows a notable degradation in the 'Weather' domain (-31.91\%), indicating potential overfitting or sensitivity to specific acoustic variations.

These results highlight the importance of model size in ASR adaptation using the \method task vector. The Base model consistently provides balanced performance across most domains, suggesting it strikes a good balance between model size and performance. In contrast, the Tiny and Small models show varying degrees of effectiveness.

Overall, the analysis demonstrates that while the \method task vector significantly improves ASR performance across different model sizes, the extent of improvement is influenced by the model's capacity. 


\subsection{Is \method task vector effective on ASR models other than Whisper?}

To validate the effectiveness of the \method task vector on other ASR models, we conduct additional experiments using the Wav2vec2-Conformer large model.\begin{table}[!ht]
\begin{adjustbox}{width=\linewidth}

\begin{tabular}{cccccc}
\toprule
\textbf{Wav2Vec2-Conformer}                                                        & \textbf{Cooking} & \textbf{Music}   & \textbf{Social}  & \textbf{Weather} & \textbf{Average}          \\ \midrule
\begin{tabular}[c]{@{}c@{}}Target Synthetic ASR\\ (Baseline)\end{tabular} & 21.26   & 17.41   & 25.84   & 16.74   & 20.31  \\ \midrule
+ \method                                                                & \textbf{18.88}   & \textbf{14.33}   & \textbf{21.48}   & \textbf{13.36}   & \textbf{17.01}   \\
Relative WER $\uparrow$                                                             & \textbf{11.21\%} & \textbf{17.66\%} & \textbf{16.87\%} & \textbf{20.22\%} & \textbf{16.25\%} \\ \bottomrule
\end{tabular}
\end{adjustbox}
\caption{\textbf{WER of \method task vector on Wav2Vec2-Conformer.} This table shows the WER and relative WER improvement across different target domains on Wav2Vec2-Conformer model before and after applying \method task vector. }
\label{tab:confomer}

\end{table}

Table~\ref{tab:confomer} presents the WER results across various target domains, including 'Cooking', 'Music', 'Social', and 'Weather', comparing the baseline model finetuned on synthetic speech with the model enhanced by the \method task vector.
The Table~\ref{tab:confomer} shows a significant reduction in WER when the \method task vector is applied. The average WER drops from 20.31 to 17.01, representing an overall relative improvement of 16.25\%.

The most notable improvement is observed in the 'social' domain, with a relative WER reduction of 16.87\%. The 'Music' domain also shows a substantial improvement of 17.66\%, indicating that the task vector successfully captures and mitigates the acoustic variability associated with music-related speech.

In the 'Cooking' and 'Weather' domains, the WER reductions are 11.21\% and 20.22\%, respectively. While the improvement in the 'Cooking' domain is more modest, it still indicates that the \method task vector enhances the model's adaptability to domain-specific acoustic characteristics.

Overall, the application of the \method task vector significantly enhances the performance of the Wav2vec2-Conformer large model across all tested domains. These results validate the effectiveness of the \method approach in bridging the gap between synthetic and real speech data, ultimately improving the robustness and versatility of ASR systems in diverse real-world scenarios.

\subsection{Can we form \method task vector from other TTS models?}

To answer \textbf{Q4}, we conducted experiments using the Whisper Small model with synthetic data generated by the Speech T5 model and XTTS model. Table~\ref{tab:t5} presents the WER results across various target domains, including 'Cooking', 'Music', 'Social', and 'Weather', comparing the baseline model finetuned on synthetic speech with the model enhanced by the \method task vector.


\begin{table}[h!]
\begin{adjustbox}{width=\linewidth}
\begin{tabular}{ccccccc}
\toprule
\textbf{TTS Model} && \textbf{Cooking} & \textbf{Music}   & \textbf{Social}  & \textbf{Weather} & \textbf{Average} \\ \midrule
\multirow{3}{*}{\textbf{Speech T5}} &Baseline (Synthetic ASR)                   & 16.94            & 16.04            & 53.34            & 16.27            & 25.65            \\ 
& + \method                                  & \textbf{16.00}   & \textbf{15.75}   & \textbf{52.95}   & \textbf{15.97}   & \textbf{25.17}   \\
& Relative WER $\uparrow$                    & \textbf{5.57\%}  & \textbf{1.77\%}  & \textbf{0.73\%}  & \textbf{1.82\%}  & \textbf{1.86\%}  \\ 

\midrule
\multirow{3}{*}{\textbf{XTTS}} & Baseline (Synthetic ASR)                   & 14.70            & \textbf{15.89}   & 23.49            & 15.52            & 17.40            \\ 
& + \method                                  & \textbf{13.51}   & 16.37            & \textbf{22.50}   & \textbf{15.06}   & \textbf{16.86}   \\
& Relative WER $\uparrow$                    & \textbf{8.11\%}  & -2.98\%          & \textbf{4.19\%}  & \textbf{2.93\%}  & \textbf{3.10\%}  \\ 
\bottomrule
\end{tabular}
\end{adjustbox}
\caption{\textbf{WER on Whisper Small with \method task vector using Speech T5 and XTTS models.} The table shows WER and relative WER improvement across different target domains.}
\label{tab:t5}
\end{table}

The results indicate that applying the \method task vector leads to a reduction in WER across most of tested domains. The average WER drops from 25.65 to 25.17 for Speech T5 and from 17.40 to 16.86 for XTTS, representing an overall relative improvement of 1.86\% and 3.10\% respectively.

The 'Cooking' domain shows the highest relative WER reduction of 5.57\% and 8.11\%, suggesting that the \method task vector effectively adapts the model to this specific domain. 

However, the improvement in the 'social' domain is relatively modest for Speech T5, with a relative WER reduction of only 0.73 \%. This could be attributed to the high baseline WER in this domain, suggesting that the synthetic data from Speech T5 might have limitations.

Overall, the application of the \method task vector to the Whisper Small model with synthetic data from different TTS models demonstrates consistent improvements in average WER, albeit with varying degrees of improvement across different domains. These results validate the flexibility and effectiveness of our approach in improving ASR models trained with synthetic data from different TTS models.


\subsection{What is the impact of the scaling factor \texorpdfstring{$\lambda$}{lambda}?}

\begin{figure}[ht]
    \begin{adjustbox}{width=\linewidth}
    \centering
    \includegraphics{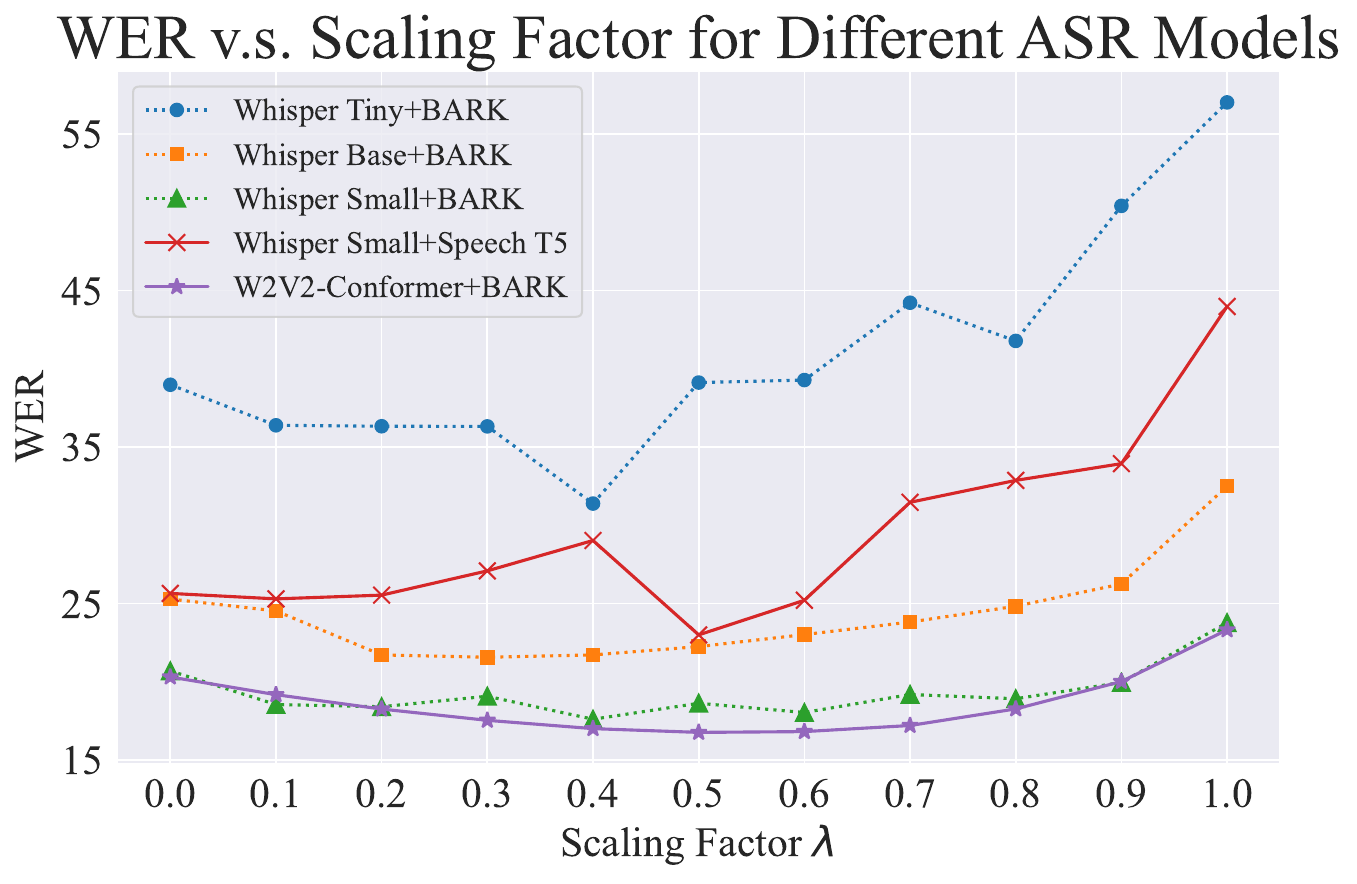}
    \end{adjustbox}
    \caption{\textbf{WER vs. Scaling Factor across Different ASR Models \& Different TTS Models } The plot shows the average WER on 'Cooking', 'Music', 'Social', and 'Weather' target domains as a function of the scaling factor $\lambda$ for various ASR models (Whisper and W2V2-conformer) and the TTS models (BARK and Speech T5) to make \method task vectors. We denote it as '\{ASR+TTS\}', such as 'Whisper Tiny+BARK' in the figure.
    The scaling factor adjusts the magnitude of the SYN2REAL task vector applied to each model.}
     \label{fig:scale}
\end{figure}

This section investigates the effect of scaling the \method task vector on the WER of different ASR models. Figure~\ref{fig:scale} illustrates the WER as a function of the scaling factor $\lambda$ for various ASR models and synthetic data, including Whisper Tiny with BARK, Whisper Base with BARK, Whisper Small with BARK, Whisper Small with Speech T5, and W2V2-Conformer with BARK. 

The scaling factor $\lambda$ adjusts the magnitude of the \method task vector applied to the ASR models. We evaluated a range of scaling factors from 0.1 to 1.0 to determine the optimal balance that minimizes WER.

The results show that different models respond variably to changes in the scaling factor. For Whisper Tiny+BARK, the curve is steeper, indicating that smaller models may be more sensitive to larger adjustments from the \method task vector. In contrast, Whisper Base+BARK maintains relatively stable WER values across different scaling factors, suggesting a more robust performance.

Notably, Whisper Small+BARK and Whisper Small~+~Speech T5 exhibit a U-shaped trend, where moderate scaling factors (around $\lambda$ = 0.3 to 0.5) yield the lowest WER. This indicates that an optimal scaling factor exists for these models, which balances the incorporation of real speech characteristics without overwhelming the model with excessive parameter adjustments. The Wav2vec2-Conformer model consistently shows lower WER values across all scaling factors, with the best performance at $\lambda$ = 0.5.

Overall, the analysis suggests that the optimal scaling factor $\lambda$ varies depending on the ASR model's architecture and size. While smaller models like Whisper Tiny+BARK may benefit from lower scaling factors, larger and more robust models like W2V2-Conformer+BARK can effectively leverage higher scaling factors. These findings highlight the importance of tuning the scaling factor to achieve the best domain adaptation performance for different ASR models.


\subsection{Do \method task vectors obtained with the same TTS have similar directions?}
\begin{figure}[ht!]
    \begin{adjustbox}{width=\linewidth}
    \centering
    \includegraphics{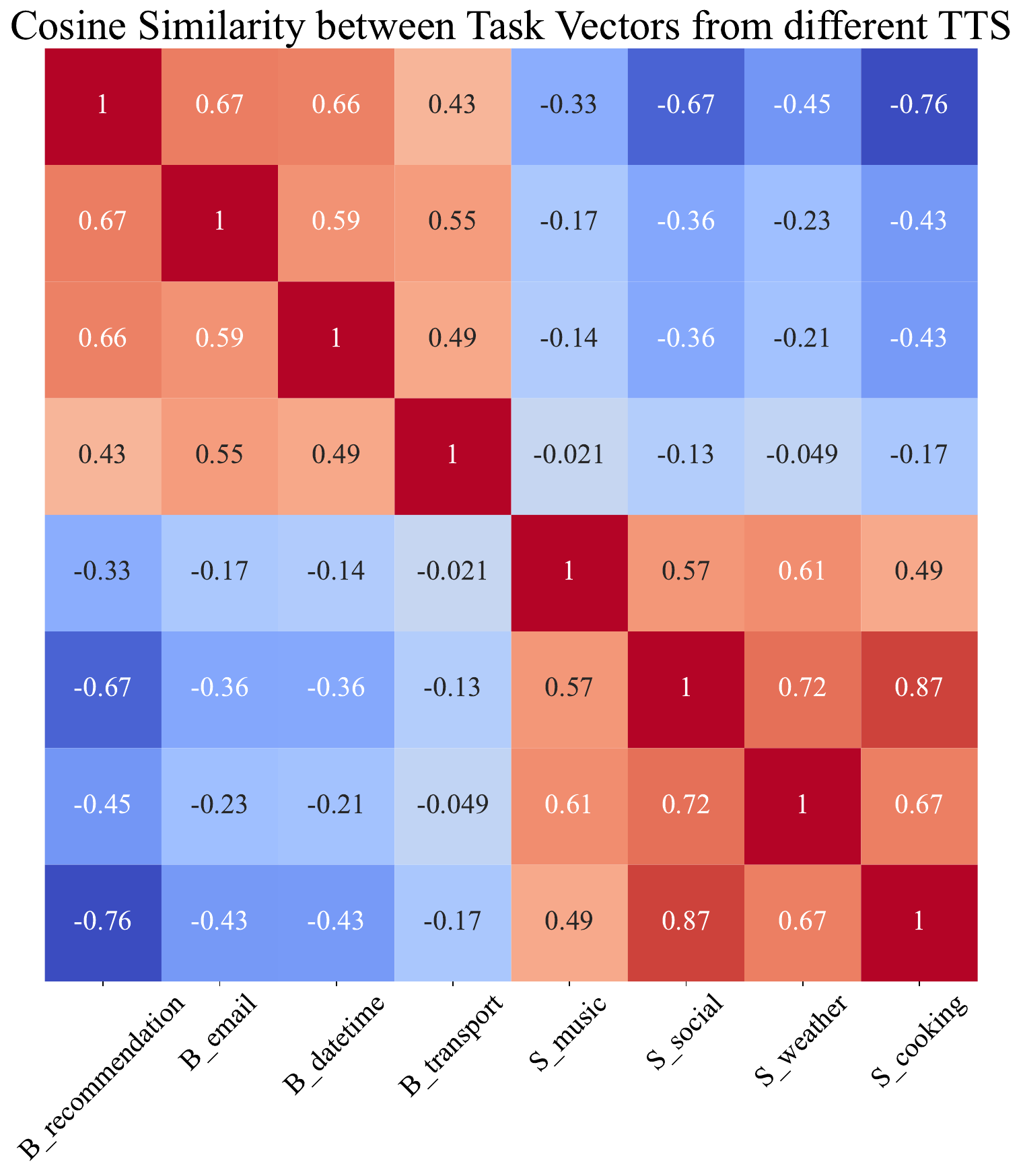}
    \end{adjustbox}
    \caption{\textbf{Cosine Similarity between task vectors  derived from Different TTS Models.} This heatmap shows the cosine similarity between task vectors generated by BARK (B\_) and Speech T5 (S\_) models. Higher similarity values between vectors from similar domains indicate effective acoustic-specific information transfer by the \method method. }
     \label{fig:heat}
\end{figure}

To further validate the \method approach, we conducted a cosine similarity analysis between \method task vectors derived by different text-to-speech (TTS) models: BARK (denoted as B\_) and Speech T5 (denoted as S\_). Figure~\ref{fig:heat} presents the cosine similarity heatmap between these \method task vectors. 

The heat map reveals that \method task vectors from similar TTS exhibit higher cosine similarity, indicating that the \method task vector effectively captures the distributional shifts between different acoustic domains. 

Moreover, the negative similarities between certain \method task vectors, such as 'B\_recommendation' and 'S\_music' (-0.67), highlight the distinct acoustic features between these TTS synthetic data, further emphasizing the effectiveness of the \method approach in distinguishing and adapting to different acoustic environments.

The overall trend observed in the heatmap supports the hypothesis that the \method task vectors not only bridge the gap between synthetic and real data but also maintain consistency within similar acoustic environments. This consistency is crucial for enhancing ASR performance across diverse target domains, as it ensures that the task vectors can generalize well to new, unseen domains.

\section{\method Task Vector given Domain Labels}
\label{sec:dis}
In this section, we explore an alternative approach to generating \method task vectors, assuming we have access to domain labels for the data in the source domains.
This approach, which we refer to as \methodd task vector, involves generating separate \method task vectors for each source domain and then combining them to enhance the adaptation of the ASR model to the target domain.


\begin{table}[h]
\begin{adjustbox}{width=\linewidth}

\begin{tabular}{cccccc}
\toprule
\textbf{\methodd}                                                                       & \textbf{Cooking} & \textbf{Music}   & \textbf{Social}  & \textbf{Weather} & \textbf{Average}           \\ \midrule
\begin{tabular}[c]{@{}c@{}}Target Synthetic ASR\\ (Baseline)\end{tabular} & 14.26            & 23.51            & 29.57            & 15.45           & 20.70   \\ \midrule
+ \methodd                                                      & 14.46   & \textbf{16.98}   & \textbf{21.13}   & \textbf{15.11}  & \textbf{16.92}   \\
Relative WER $\uparrow$                                                             & -1.40\% & \textbf{27.78\%} & \textbf{28.54\%} & \textbf{2.20\%} & \textbf{18.25\%} \\ \bottomrule
\end{tabular}
\end{adjustbox}
\caption{\textbf{WER Performance on Whisper Small Model with \methodd task vectors.} This table compares the word error rate (WER) of the baseline ASR model fine-tuned on synthetic speech data with the WER of the model enhanced with the \methodd task vectors across four target domains: 'Cooking', 'Music', 'Social', and 'Weather'. }
\label{tab:dis}
\end{table}
\subsection{Performance of \methodd task vector}

To evaluate the effectiveness of the \methodd task vector, we conducted experiments using the Whisper Small model with synthetic speech generated by the BARK TTS model. The experiments were carried out on four target domains: 'Cooking', 'Music', 'Social', and 'Weather,' using 17 source domains to create the \methodd task vectors. The results are presented in Table~\ref{tab:dis}.
The results indicate that the \methodd task vector provides significant improvements in WER for several target domains compared to the baseline method. The average WER is reduced from 20.70 to 16.92, representing an overall relative improvement of 18.25\%. This highlights the effectiveness of using domain-specific task vectors to capture detailed acoustic characteristics, leading to enhanced model adaptation.


Comparing the \methodd task vector to the original \method task vector, we find that \methodd task vector generally outperforms the original approach. The detailed domain-specific information captured by the distinct task vectors enhances model adaptation in most domains. However, in real-world scenarios, we often do not have access to labels finer-grained domain labels.

\subsection{Impact of the number of domains on the performance of \methodd task vector}
\begin{figure}[tp!]
    \begin{adjustbox}{width=\linewidth}
    \centering
    \includegraphics{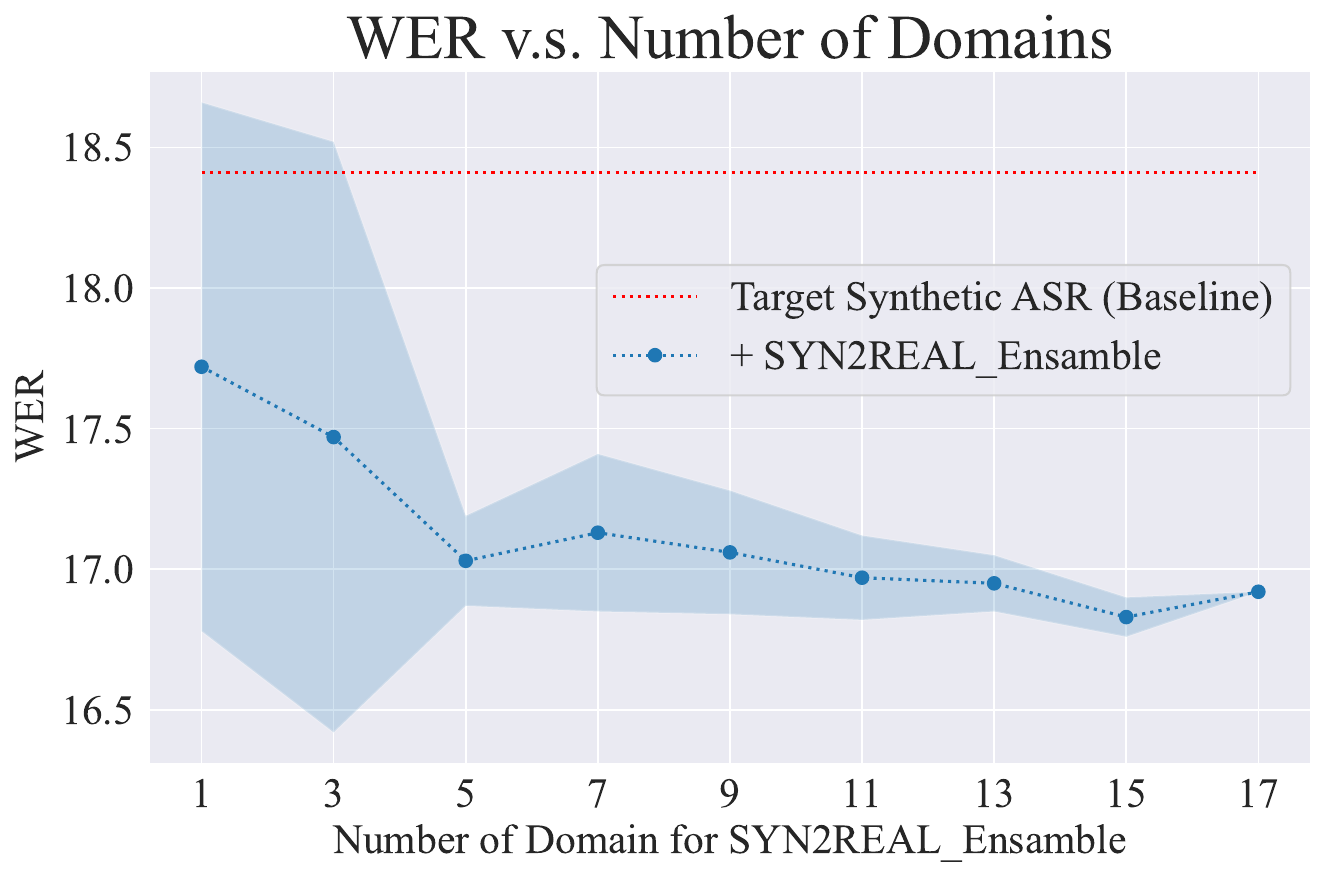}
    \end{adjustbox}
    \caption{\textbf{WER vs. Number of Source Domains for \methodd task vector.} This plot shows the word error rate (WER) of the Whisper small model and the number of source domains used to generate the \methodd task vector with BARK model. The x-axis represents the number of source domains, and the y-axis represents the WER on average of the four domains ('Cooking', 'Music', 'Social', and 'Weather').}
     \label{fig:domain}
\end{figure}

Figure~\ref{fig:domain} shows the average WER across four target domains ('Cooking', 'Music', 'Social', and 'Weather') when we use different numbers of source domain data to generate \methodd task vectors.\looseness=-1

The results indicate that increasing the number of source domains generally improves ASR performance. The WER decreases from 17.8 to 16.8 as the number of source domains increases from 1 to 17. Notably, significant improvements are observed within the incorporation of the first 5 source domains. This trend suggests that incorporating more source domains helps capture diverse acoustic characteristics, leading to better domain adaptation.

\section{Conclusion}



This paper introduces \method task vector to address mismatches between synthetic and real speech data. Future work will refine this approach and extend its application to other types of tasks and data such as visual data, contributing to more reliable speech and vision recognition systems.
Experiments showed significant WER reductions, averaging 10.03\% across 18 domains. 
We test various models, including Whisper Small, Whisper Base, Whisper Tiny, and Wav2Vec2-Conformer, with \method task vector showing robust performance.
We also explore \methodd approach further enhanced domain-specific adaptation but required domain labels. 
Overall, \method task vector is a promising solution for improving ASR models with synthetic data.

\section{Limitations}
\paragraph{Domain-Specific Performance Variations} While the \method task vector shows significant improvements in many target domains, certain domains, such as 'Cooking' and 'Weather,' exhibit marginal improvements or slight degradation in word error rate (WER). This suggests that the task vector's effectiveness may vary based on the specific characteristics of different domains, indicating a need for further domain-specific fine-tuning and adjustments.

\paragraph{Scaling Factor Sensitivity} The performance of the \method-enhanced models is sensitive to the scaling factor $\lambda$. Finding the optimal scaling factor requires careful tuning, and the best value can vary between different ASR models and target domains. This adds a layer of complexity to the implementation and may limit the approach's generalizability without additional adaptive scaling strategies.

\paragraph{Synthetic Data Quality} The approach relies heavily on the quality of synthetic speech data generated by TTS systems. Variations in the quality and acoustic properties of synthetic data across different TTS systems can impact the effectiveness of the \method task vector. Ensuring consistent quality in synthetic data is crucial for achieving robust domain adaptation.

\paragraph{Model-Specific Dependencies} The observed improvements are model-dependent, with larger models like Wav2Vec2-Conformer showing more substantial gains compared to smaller models like Whisper Tiny. This indicates that the \method task vector's effectiveness might be influenced by the underlying model architecture and size, potentially limiting its applicability to a wider range of ASR models without further optimization.

\section{Acknowledgements}
We specifically thank Ting-Yao Hu and Dianna Yee for all the insightful discussions and constructive suggestions for this work.

This work was supported in part by the National Science and Technology Council under Grants NSTC 112-2634-F-002-006, MOST 110-2222-E-002-014-MY3, NSTC 113-2222-E-002-004-MY3, NSTC 113-2634-F-002-001-MBK.


\bibliography{custom}

\appendix



\end{document}